# Exchange field enhanced upper critical field of the superconductivity in compressed antiferromagnetic EuTe$_2$

Hualei Sun[1], Liang Qiu[1], Yifeng Han[2], Yunwei Zhang[1], Weiliang Wang[3], Chaoxin Huang[1], Naitian Liu[1], Mengwu Huo[1], Lisi Li[1], Hui Liu[1], Zengjia Liu[1], Peng Cheng[4], Hongxia Zhang[4], Hongliang Wang[5], Lijie Hao[5], Man-Rong Li[2], Dao-Xin Yao[1], Yusheng Hou[1], Pengcheng Dai[4], and Meng Wang[1,*]

[1]*Center for Neutron Science and Technology, Guangdong Provincial Key Laboratory of Magnetoelectric Physics and Devices, School of Physics, Sun Yat-Sen University, Guangzhou, Guangdong 510275, China*

[2]*Key Laboratory of Bioinorganic and Synthetic Chemistry of Ministry of Education, School of Chemistry, Sun Yat-Sen University, Guangzhou, Guangdong 510275, China*

[3]*School of Physics, Guangdong Province Key Laboratory of Display Material and Technology, Sun Yat-Sen University, Guangzhou, Guangdong 510275, China*

[4]*Laboratory for Neutron Scattering and Beijing Key Laboratory of Optoelectronic Functional Materials and MicroNano Devices, Department of Physics, Renmin University of China, Beijing 100872, China*

[5]*China Institute of Atomic Energy, PO Box-275-30, Beijing 102413, China*

[6]*Department of Physics and Astronomy, Rice Center for Quantum Materials, Rice University, Houston, TX, USA*

\* [wangmeng5@mail.sysu.edu.cn](wangmeng5@mail.sysu.edu.cn)

## Abstract

We report high pressure studies on the *C*-type antiferromagnetic semiconductor EuTe$_2$ up to 36.0 GPa. A structural transition from the *I4/mcm* to *C2/m* space group is identified at ~16 GPa. Superconductivity is discovered above ~5 GPa in both the *I4/mcm* and *C2/m* space groups. In the low-pressure phase (< 16 GPa), the antiferromagnetic transition temperature is enhanced with increasing pressure due to the enhanced magnetic exchange interactions. Magnetoresistance measurements indicate an interplay between the local moments of Eu$^{2+}$ and the conduction electrons of Te 5*p* orbits. The upper critical field of the superconductivity is well above the Pauli limit. Across the structural transition to the high-pressure phase (> 16 GPa), EuTe$_2$ becomes nonmagnetic and the superconducting transition temperature evolves smoothly with the upper critical field below the Pauli limit. Therefore, the high upper critical field of EuTe$_2$ in the low-pressure phase is due to the exchange field compensation effect of the Eu magnetic order and the superconductivity in both structures may arise in the framework of the BCS theory.

## I. INTRODUCTION

Superconductivity in conventional Bardeen-Cooper-Schrieffer (BCS) superconductors arises from electron-lattice interaction without the involvement of magnetism[1]. Below the superconducting (SC) transition temperature, electrons form coherent spin singlet cooper pairs that can be suppressed by a Pauli limited magnetic field. In contrast, one of the hallmarks of unconventional superconductivity is the interplay between magnetism and superconductivity. For example, superconductivity in copper oxide and iron-based high temperature superconductors occurs near long-range magnetic order where the 3*d* electrons of the transition metals across the Fermi level contribute to both the magnetic correlations and superconductivity[2–6]. In most unconventional superconductors, electron pairing forms spin singlet and upper critical field needed to suppress superconductivity is also Pauli limited[2-6]. For unconventional superconductivity with upper critical field exceeding the Pauli limit, such as recently discovered UTe$_2$[7,8], electron pairing is believed to be spin triplets instead of singlets. In both spin singlet and spin triplet superconductors, magnetic fluctuations play an important role in the formation of cooper pairs as evidenced by the neutron spin resonance from inelastic neutron scattering spectrum.[6,9]

Although the mechanism of superconductivity for conventional and unconventional superconductors may be fundamentally different, both superconductors can also host local moment magnetic ions not directly associated with SC layers. For example, in a class of iron-based superconductors consisting of Eu$^{2+}$, the 4$f$ electrons with spin $S = 7/2$ could form an antiferromagnetic (AFM) or ferromagnetic (FM) sublattice coexisting and interacting with the magnetic sublattice of Fe[10,11]. However, the localized magnetism of Eu$^{2+}$ does not interplay with the superconductivity seriously. For BCS superconductors such as $R$Ni$_2$B$_2$C series ($R$= Y, Er, Ho etc)[9], the interplay between AFM (FM) order of the rare earth layers and superconductivity can dramatically affect the physical properties of the system including the upper critical field needed to suppress superconductivity. In 1962, Jaccarino and Peter proposed that the effective exchange field ($H_J$) from a FM rare earth metal impressed on the conduction electrons via the exchange interaction with the rare earth spin $S$ could oppose or cancellate the external magnetic field ($H$) resulting in an ultra-high upper critical field ($H_{c2}$) superconductivity contrasting the expectation from the BCS theory[12]. The exchange field compensation effect indeed be observed in Eu-containing metallic compounds where the 4$f$ electrons of Eu$^{2+}$ form a large exchange field $H_J$ that is opposite to that of $H$[13]. While the compensation effect is rare because magnetism normally suppresses superconductivity for the BCS superconductors.

Previously, our group reported an antiferromagnetically colossal angular magnetoresistance EuTe$_2$ with a Néel temperature of 11.4 K from the Eu local moment and a thermal-activation gap of 16.24 meV at atmospheric pressure[14]. The magnetic field drives polarization of the local moments of Eu$^{2+}$ and results reconstruction of the Te 5$p$ orbitals through the exchange couplings and the space-time inversion symmetry-broken[14,15]. While density function theory (DFT) calculations could not distinguish the specific AFM from the $A$-type and $C$-type orders[14]. Pressure is a pure and effective way to tune lattice parameters and overlapping of the electronic orbitals. Superconductivity has been realized in the Te-containing compounds CrSiTe$_3$ and WTe$_2$ under pressure[16,17]. In EuIn$_2$As$_2$ and EuSn$_2$As$_2$, the Néel temperatures are enhanced under pressure due to the increasement of magnetic exchange couplings[18,19]. A structural transition accompanying by a possible valent state transition from Eu$^{2+}$ to Eu$^{3+}$ is observed. It is expected that pressure can induce the semiconductor to metal transition and enhance the magnetic correlations of the local moments of Eu$^{2+}$. Very recently, a high-pressure study on EuTe$_2$ up to 12.0 GPa indeed reveal superconductivity and suggests the SC pairing mechanism is exotic[20].

In this work, we present comprehensive experimental and theoretical investigations on EuTe$_2$ under pressure up to 36.0 GPa. Neutron diffraction measurements demonstrate EuTe$_2$ exhibits a $C$-type AFM order at low temperature. A pressure induced structural transition at ~16 GPa is discovered. In the low-pressure (LP) phase, the $C$-type AFM transition temperature $T_N$ increases due to the enhancement of the magnetic exchange interactions of the compressed lattice. The thermal activation gap $E_a$ is closed progressively, and superconductivity emerges above 5.0 GPa. The SC transition temperature $T_c$s spanning between 5~7 K in the pressure range of 5~27 GPa is irrespective to the structural transition and magnetism. While the upper critical field $H_{c2}$ for the superconductivity of the AFM LP phase is significantly larger than that of the superconductivity of the nonmagnetic (NM) high pressure (HP) phase.

The $H_{c2}$ is obviously affected by the microscopic magnetic order of the Eu sublattice, possibly due to the Jaccarino and Peter mechanism. The highest $\rho_0 H_{c2}$ is estimated to be 21.6 T for the spin flipped state at 7.0 GPa. The ultra-high $H_c$ could be understood by the compensation effect of the exchange field of $Eu^{2+}$. Our results therefore establish the pressure-temperature phase diagram of $EuTe_2$, and demonstrate the interesting interplay between Eu magnetic order, superconductivity, and pressure-induced structural lattice distortion.

## II. RESULTS
### A. High pressure structure

Figure 1 displays the *in situ* high pressure synchrotron powder x-ray diffraction (XRD) patterns of $EuTe_2$ up to 36.0 GPa at room temperature and the refined crystal structures below 15.9 GPa and above 17.9 GPa, defined as the LP phase and HP phase, respectively. The LP phase can be indexed by the trigonal *I4/mcm* space group (No. 140), identical to the ambient pressure crystal structure. The divalent europium is coordinated by eight nearest-neighbor tellurium ions[14]. The edge-sharing octagonal units form the layers of the tetragonal crystal structure as shown in Fig. 1(c).

In terms of the diffraction peaks changed under pressure, an obvious structural phase transition between 15.9 and 17.9 GPa could be identified. We conducted an extensive search on the HP structure of $EuTe_2$ in the pressure range of $0 - 25$ GPa via the *CALYPSO* method[21–23]. The monoclinic *C2/m* (No. 12) structure turns to be a possible candidate of the HP phase at 17.9 GPa. Thus, we refined the experimental XRD pattern at 17.9 GPa by the Rietveld method through the *TOPAS*-Academic software[24]. The *C2/m* structure matches the XRD pattern of the HP phase well (see supplementary). Figure 1(b) shows the structure of the HP phase. The europium ions retain the eight-coordination but there is a significant deformation of the octagonal unit. This coordination unit exists in the compounds of $Eu_3S_4$ at atmospheric pressure, which confirms it is a stable coordination structure for europium chalcogenide[25]. Sulfur and tellurium both are chalcogenides, but sulfur has a smaller ionic radius than tellurium which is equivalent to pressurizing tellurium.

For the HP phase, slip occurs between the adjacent layers compared with the LP phase. As shown in Figs. 1(b) and 1(c), the unit cell volume decreases sharply at the pressure-induced structural transition from 317.157(9) $Å^3$ at 15.9 GPa to 262.524(3) $Å^3$ at 17.9 GPa, which may be accompanied by the valent state transition from $Eu^{2+}$ to nonmagnetic $Eu^{3+}$. The diffraction peaks from Te impurity and structural transitions of Te can be observed in Fig. 1(a)[26,27]. Tellurium as the flux in the single crystal growth is hard to be eliminated. The refined XRD patterns and structural parameters for 9.7 and 17.9 GPa are shown in the Supplementary figures.

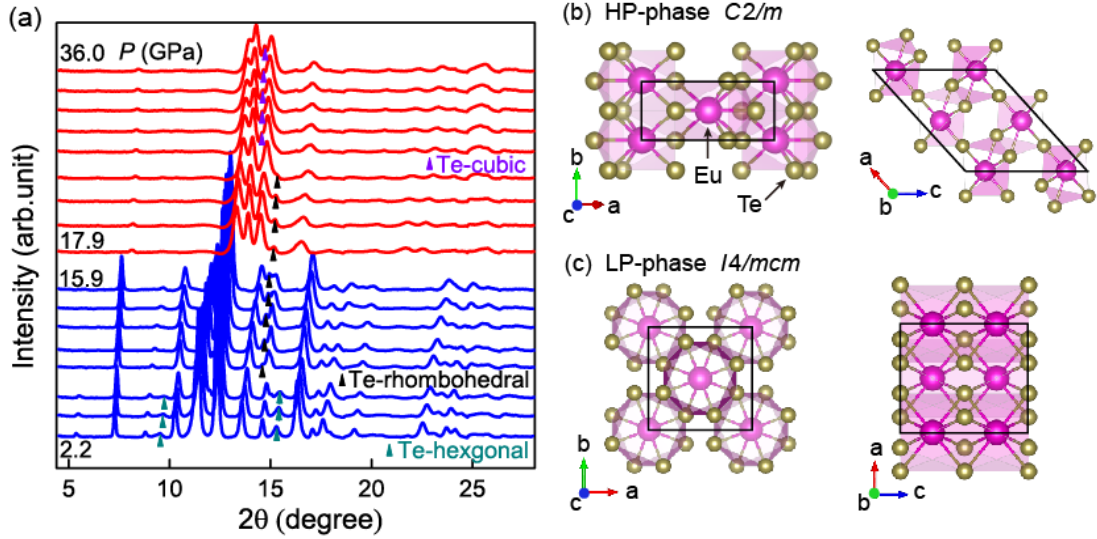

**Figure 1** (a) High-pressure XRD patterns of EuTe$_2$ from 2.2 to 36.0 GPa with an x-ray wavelength of 0.6199 Å. The XRD patterns of the LP phase are in blue and the patterns of HP phase are in red. Peaks from Te impurity are marked by the triangles. Tellurium undergoes two structural phase transitions within the measured pressure. (b) Crystal structures of EuTe$_2$ of the HP phase and (c) the LP phase.

### B. High pressure electrical and magnetic properties

To investigate the electrical properties of EuTe$_2$ under pressure, we performed electrical transport measurements below 27.7 GPa. Figure 2(a) shows temperature dependence of the resistance at various pressures, revealing semiconducting to metallic and to SC transitions. Resistance as a function of pressure for selected temperatures is presented in Fig. 2(b). The magnitude of the resistance decreases as pressure increasing. The upturn in resistance at low pressure may be attributed to the scattering of conduction electrons by local moments of the Eu$^{2+}$ ions. An abrupt drop in resistance appears between 14.7 and 16.2 GPa, consistent with the structural transition between 15.9 and 17.9 GPa. Thus, the structural transition pressure should occur at ~16.0(2) GPa. The resistance above 50 K in Fig. 2(a) is fitted to the thermal activation-energy model $\rho(T) = \rho_0 \exp(E_a/k_B T)$, where $\rho_0$ is a prefactor, $E_a$ is the thermal activation gap, and $k_B$ is the Boltzmann constant. The gap of 16.24 meV for EuTe$_2$ at ambient pressure is gradually closed by pressure, as shown in Fig. 2(c). The evolution of the carriers against pressure is also investigated by the Hall resistance measurements. The Hall coefficient remains positive, revealing that the majority carriers are holes (see supplementary). The determined density of holes shows an abrupt enhancement across the structural transition like the observation in EuSn$_2$As$_2$[19].

To elucidate the magnetic state of the HP phase, we conducted systematic magnetoresistance (MR) measurements against temperature and pressure. At ambient pressure, EuTe$_2$ shows colossal negative MR resulting from the splitting of the Te 5$p$ orbitals induced by the exchange field of localized Eu$^{2+}$ spins. Under pressure, the semiconducting gap is decreased, the resistance without magnetic field becomes much smaller and the MR is suppressed accordingly. The integrated $MR$s (defined as $MR = (\rho_H - \rho_0)/\rho_0$) over the magnetic fields from –10 to 10 T as presented in Fig. 2(d) decrease as pressure and temperature, diminishing gradually above 16.2 GPa. The absence of MR suggests that the HP phase is paramagnetic.

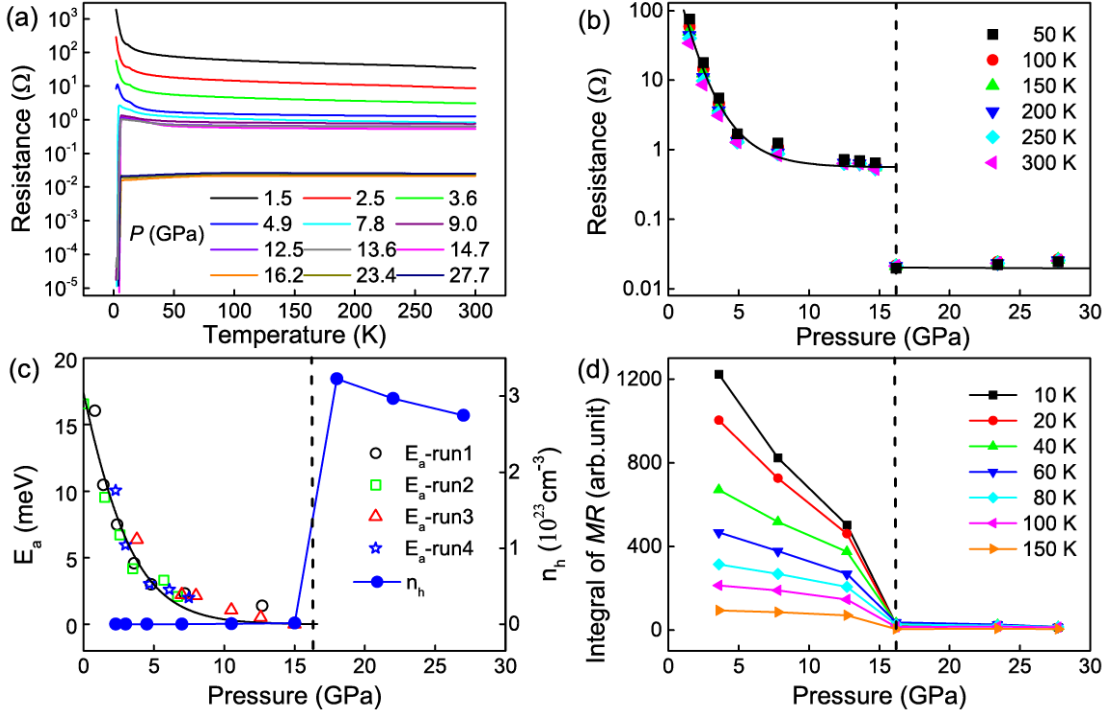

**Figure 2** (a) Temperature dependence of the resistance upon pressures up to 27.7 GPa. (b) Pressure dependence of the resistance at various temperatures up to 300 K shown on a logarithm scale. (c) Thermal-activation gaps derived from fittings of the resistance curves within the temperature range from 60 to 300 K using $\rho(T) = \rho_0 \exp(E_a/k_B T)$. Different shapes of data points are measured on different samples. The black solid line is a guide to the eyes. On the right scale, carrier densities as a function of pressure. (d) Integrals of $MR$ over a magnetic field from −10 to 10 T as a function of pressure for selected temperatures from 10 to 150 K. The dashed lines in (b-d) at 16.0 GPa mark the pressure of the structural transition.

Figure 3 shows the resistance of Fig. 2(a) as a colormap on a logarithm scale. The $T_N$ of the AFM transition and $T_c$ of the SC transition under pressure could be identified from the resistance (see supplementary). Upon increasing pressure, the derived $T_N$s increase from 11.4 K at ambient pressure to 16.7 K at 8.0 GPa. The superconductivity appears at 4.8 GPa with a $T_c$ of 4.1 K, defined by the intersection of the tangent to the resistance curve during the transition process and the straight-line of the normal state above the SC transition. The $T_c$ reaches a maximum of 6.1 K at 7.0 GPa and decreases smoothly afterwards across the structural transition, indicating that the AFM order and spin fluctuations of $Eu^{2+}$ have not contributed to the cooper pairing mechanism directly.

At ambient pressure and low temperature, the calculated energy difference between the *A*-type AFM and the *C*-type AFM is almost neglectable (about 1.5 meV/Eu)[14]. Neutron diffraction measurements were employed to distinguish the two magnetic structures. Although the neutron absorption from Eu atoms is serious, the magnetic reflections associated with the *C*-type AFM are observed unambiguously (see supplementary). To understand the underlying mechanism for the enhanced $T_N$ in compressed $EuTe_2$, we investigate its exchange couplings based on the following spin model:

$$H = \sum_{\langle ij \rangle} J_{ij} \mathbf{S}_i \cdot \mathbf{S}_j + A \sum_i \left(S_i^z\right)^2 \quad (1).$$

Considering the small gap of $EuTe_2$, six nearest neighbor (NN) Heisenberg exchange couplings are considered. In Eq. (1), $A$ is the single-ion magnetic anisotropy parameter. For EuTe$_2$ at ambient pressure, our DFT calculations show that it exhibits a $C$-type AFM ground with a small gap of 18 meV and out-of-plane magnetic easy axis, consistent with our neutron scattering measurements. Our Monte Carlo simulations reveal the $T_N$ is 13.17 K close to the previous studies[14]. Under pressure, both the DFT calculations and Monte Carlo simulations indicate that the ground state is also the $C$-type AFM. Four NN exchange couplings are strengthened obviously regard of ferromagnetic ($J < 0$) or antiferromagnetic ($J > 0$) terms (see supplementary). This is understandable because the distances between the Eu$^{2+}$ ions decrease under pressure. Correspondingly, the calculated $T_N$s increase from 13.17 to 21.21 K at 11.8 GPa as shown in Fig. 3, consistent with our experimental observations below 10 GPa.

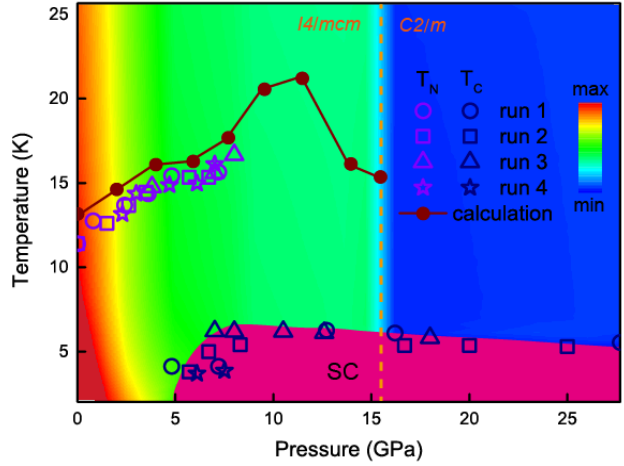

**Figure 3** A phase diagram of the AFM transition temperature $T_N$ and SC transition temperature $T_c$ against pressure. The filled circles are calculated $T_N$s. The color represents different resistance on a logarithm scale. The red color represents high resistance and the blue one means low resistance. Different shapes of data points are obtained from different measurements.

### C. Pressure induced superconductivity

Superconductivity emerges in both the LP and HP phases, which have distinct structures and magnetic ground states. Through DFT calculation, the localized Eu 4$f$ electrons reside ~1.25 eV below the Fermi level for the LP phase, while the Te 5$p$ electrons crossing the Fermi level involve SC cooper pairing (see supplementary). To explore the role of the local moments of Eu$^{2+}$ in superconductivity, we conducted resistance measurements at 7.0 GPa. Figures 4(a) shows the resistance against magnetic field for selected temperatures. Figure 4(b) displays resistance measured against temperature at various magnetic fields. The $T_c$s are lowered by magnetic field as expected. Surprisingly, the superconductivity persists up to 14 T which is the highest magnetic field in our measurements. The $T_c$s determined from the resistance in Fig. 4(b) and resistance shown as a colormap on a linear scale are displayed in Fig. 4(c). We find the $H_{c2} - T_c$ relation does not follow a simple Ginzburg-Landau (GL) formula, $\mu_0 H_{c2}(T) = \mu_0 H_{c2}(0)[1 - (\frac{T}{T_c})^2]$. The experimentally determined $T_c$s against magnetic field could be separated into three segments.

We note the $C$-type AFM structure of EuTe$_2$ at ambient pressure undergoes a spin flop transition at ~3.0 T and a spin flip transition at ~8.0 T and 1 K[14]. The upper critical field $H_{c2}$ may be affected by the exchange field $H_J$ produced by the local moments of Eu$^{2+}$. In this case, the net magnetic field $H_T$ acting on the conduction electrons is $H_T = H_{c2} - |H_J|$.[28] The magnetic fields for spin flop and spin flip transitions below the $T_N$ of 15.6 K at 7.0 GPa are determined from the resistance in Fig. 4(a). A hump on resistance below the $T_N$ and above the $T_c$ could be attributed to the spin transitions (see supplementary).

The two magnetic fields corresponding to the spin flop and spin flip transitions at ~5 K and 7.0 GPa are 5.5 and 12.5 T, respectively. The increased values compared with that at ambient pressure are proportional to the increase of the $T_N$. The magnetic phase diagram is consistent with the three segments of the $H_{c2} - T_c$ relation. Thus, the $H_{c2}$s are fitted for the AFM, spin flop, and spin flipped states to the GL formula, resulting in the upper critical fields of 10.1, 16.2, and 21.6 T, respectively. The $H_{c2}$s for the spin flop and spin flipped states are well above the Pauli limit of $\rho_0 H_{c2} = 1.84 \times T_c = 11.2$ T, where $T_c$ is 6.1 K at 7.0 GPa and zero field[29]. As the Jaccarino-Peter mechanism, AFM spins do not contribute to exchange field. The spin flipped state with fully polarized spins of $Eu^{2+}$ has the maximum $H_J$. If the sign of the coupling between the local spins and conduction electron spins is negative, the measured $H_{c2}$ should be larger than the Pauli limit[12]. As a comparison, we show the resistance under various magnetic fields and $H_{c2}$s at 18.0 GPa in Fig. 4(d). The colormap of resistance suggests that the HP phase is nonmagnetic. The $H_{c2}$ can be described by a single GL formula with the $T_c = 5.5$ K and $\rho_0 H_{c2} = 6.15$ T within the Pauli limit of 10.12 T.

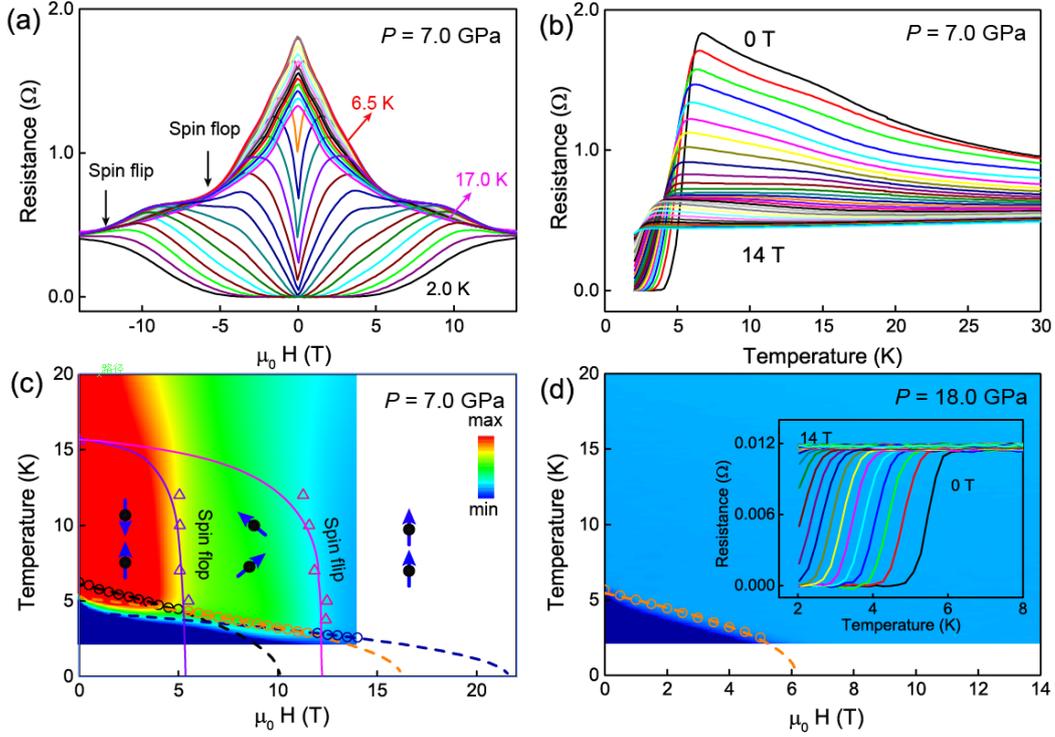

**Figure 4** (a) Magnetic field dependence of the resistance at 7.0 GPa and selected temperatures from 2.0 to 17.0 K. (b) Resistance from 2.0 to 30.0 K in different magnetic fields from 0 to 14.0 T measured every 0.5 T at 7.0 GPa. (c) Temperature-field phase diagram with the AFM ordering temperature, spin flop and spin flip transitions, and SC transition. The solid lines are guide of the magnetic transitions. The dashed lines are the GL formula fittings to the segments of the SC transition temperatures. The fitted $H_{c2}$ are 10.1, 16.2, and 21.6 T, respectively. The color represents the value of the resistance of (b) on a linear scale. The magnetic fields of spin flop and spin flip transitions are obtained from the magnetic field dependence of resistance in (a) (see supplementary). (d) Phase diagram of the superconductivity against magnetic field at 18.0 GPa. The inset is the resistance curves measured on selected magnetic fields. The dashed line is a fitting to the Ginzburg-Landau formula.

## III. DISCUSSION AND SUMMARY

Tellurium is a superconductor with a maximum $T_c$ of ~2 K and $H_{c2}$ below 0.1 T under pressure[30]. In some compounds consisting of Te, the $H_{c2}$s could achieve the magnitude of several tesla, such as $CrSiTe_3$,

WTe$_2$, HfTe$_5$, Bi$_2$Te$_3$, and CsBi$_4$Te$_6$, where the normal states are not magnetically ordered[17,16, 31–33]. The $H_{c2}$s for these compounds are still below the Pauli limit. UTe$_2$ superconducts below 1.6 K at ambient pressure with an ultra-high $\mu_0 H_{c2} > 45$ T[34]. The 5$f$ electrons of uranium cross the Fermi level and contribute to the magnetism and superconductivity directly, resulting in the heavy Fermi property and possible triplet pairing mechanism[35,36]. For iron-based superconductors containing AFM or FM Eu$^{2+}$, the high upper critical fields are governed by iron pnictide or iron chalcogenide layers. The 4$f$ electrons of Eu$^{2+}$ are below the Fermi level and do not involve to the superconductivity. The contribution of the exchange field from FM Eu$^{2+}$ to the $H_{c2}$ through the compensation effect is not obvious. EuTe$_2$ with a maximum $T_c$ of 6.1 K comparable with the other Te-containing materials and $H_{c2}$s within the Pauli limit in the AFM state of the LP phase and the nonmagnetic HP phase. In the spin flop and spin flipped states, the net polarization of the moments will couple with the conduction electrons and result in an exchange field $H_J$ with an opposite sign to the external applied field $H$. The net magnetic field acting on the conduction electrons, $H_T$, is $H_T = H - |H_J|$. When $H_T > H_P$, where $H_P$ is the paramagnetic limiting field that determines the SC behavior, the external applied field $H$, that is the experimental determined $H_{c2}$, drives EuTe$_2$ to the normal state. Due to the compensation effect, the $H_{c2}$ could be much larger than the Pauli limit of the BCS theory[12,29]. For Eu$_{0.75}$Sn$_{0.25}$Mo$_6$S$_{7.2}$S$_{0.8}$, an applied magnetic field can progressively tune the compound from SC to normal, to SC again, and finally back to normal state below 1 K[13]. To estimate the $H_p$ and $H_J$ for Eu$_{0.75}$Sn$_{0.25}$Mo$_6$S$_{7.2}$S$_{0.8}$ for the second SC phase with the lower and upper $\mu_0 H_{c2}$s of 4 and 22 T, we have the constrains of: (i) 4 T$-\mu_0 H_J = -\mu_0 H_P$, and (ii) 22 T$-\mu_0 H_J = \mu_0 H_P$. The $\mu_0 H_J$ and $\mu_0 H_P$ with values of 13 and 9 T could be derived, respectively. If we assume the $H_J$s are comparable in EuTe$_2$, then the upper critical field for the spin flipped state in the LP phase of EuTe$_2$ should be $10.1 + 13 \approx 23$ T, close to the fitted $\mu_0 H_{c2}$ of 21.6 T using the GL formula.

In summary, we have studied the structural and electronical transport properties of EuTe$_2$ under pressure. EuTe$_2$ shows a SC transition above 5 GPa with a maximum $T_c$ of 6.1 K at 7.0 GPa and a structural transition at 16 GPa. The transition temperature of the $C$-type AFM that is determined from neutron diffraction is enhanced in compressed EuTe$_2$ due to the increase of the magnetic exchange interactions. In the low pressure phase, superconductivity coexists with the AFM, spin flop, and spin flipped states. However, the electronic states of Eu$^{2+}$ are well below the Fermi level and do not involve cooper pairing directly. The local moments of Eu$^{2+}$ in the spin flop and spin flipped states produce an effective magnetic field, called the exchange field, compensating with the external field and resulting in an ultra-high upper critical field that is larger than the Pauli limit. The HP phase is nonmagnetic and the $H_{c2} - T_c$ relation could be described by the Ginzburg-Landau formula with the $H_{c2}$ within the Pauli limit. Our results establish that EuTe$_2$ is a pressure-induced superconductor with the Jaccarino-Peter mechanism.

## IV. Methods

**Single-crystal growth and neutron diffraction**

Bulk single crystals of EuTe$_2$ were grown by the self-flux method as we previously reported[14]. The shiny black single crystals of EuTe$_2$ were separated from the Te flux at 450°C. The structure of EuTe$_2$ was confirmed by single crystal XRD.

The powder neutron diffraction experiments were carried out on Xingzhi triple-axis spectrometer at the China Advanced Research Reactor (CARR)[37]. Powder samples were stuck on an aluminum foil uniformly with a hydrogen free glue to reduce the absorption of Eu, then sealed in a cylindrical vanadium container and loaded into a closed cycle refrigerator that regulates the sample temperature from 3.5 to

300 K. A neutron velocity selector was used upstream to cleanly remove higher order neutrons for the incident neutron energy fixed at 16 meV.

**High-pressure XRD**

The *in situ* high pressure synchrotron powder XRD patterns of EuTe$_2$ were collected at 300 K with an x-ray wavelength of 0.6199 Å on Beijing Synchrotron Radiation Facility, Institute of High Energy Physics, Chinese Academy of Sciences (BSRF, IHEP, CAS). A symmetric diamond anvil cell (DAC) with a pair of 300-μm-diameter culets was used. A sample chamber with a diameter of 120 μm was drilled by laser in a pre-indented steel gasket. The EuTe$_2$ single crystals were ground into fine powders and compressed into a pellet with an 80-μm diameter and 20-μm thickness. The pellet was loaded into the middle of the sample chamber and silicone oil was used as a pressure transmitting medium. A ruby sphere was also loaded into the sample chamber and pressure was determined by measuring the shift of its fluorescence wavelength. The data were initially processed using *Dioptas*[38] (with a CeO$_2$ calibration) and the subsequent Rietveld refinements were managed using *TOPAS*-Academic.[24]

**High-pressure magnetic and electrical property measurements**

Magnetic and electrical measurements were taken on a physical property measurement system (PPMS, Quantum Design). High-pressure electrical transport measurements of EuTe$_2$ single crystals were carried out using a miniature DAC made from a Be–Cu alloy on a PPMS. Diamond anvils with a 400-μm culet were used, and the corresponding sample chamber (with a diameter of 150-μm) was made in an insulating gasket achieved by cubic boron nitride and epoxy mixture. NaCl powders were employed as the pressure-transmitting medium, providing a quasi-hydrostatic environment. The pressure was also calibrated by measuring the shift of the fluorescence wavelength of the ruby sphere, which was loaded in the sample chamber. The standard four-probe technique was adopted for these measurements.

**First-principles calculations**

Our structure searching simulations are performed by the swarm-intelligence based *CALYPSO* (Crystal structure AnaLYsis by Particle Swarm Optimization) method, which enables global minimization of energy surfaces by merging *ab initio* total-energy calculations. The structure searching was carried out at pressures of 5, 15, and 25 GPa which covers the experimental pressure range. The simulation cell sizes of 1– 4 formula units were set. The underlying *ab initio* structural relaxations were carried out using density functional theory within the Perdew–Burke–Ernzerhof (PBE) exchange-correlation[39] as implemented in the Vienna *ab initio* Simulation Package (*VASP*) code[40,41].

DFT calculations are performed using the *VASP* at the level of the generalized gradient approximation.[39,42] We adopted the projector augmented wave pseudopotentials and a plane-wave cutoff energy of 500 eV.[40] The experimentally measured lattice constants are used in our calculations and the positions of all atoms are fully relaxed until the force on each atom is less than 0.01 eV/Å. We use $U = 4.4$ eV for Eu$^{2+}$ ions in view of the strong correlation among $f$ electrons. The $T_N$ of the pressurized EuTe$_2$ is obtained through parallel tempering Monte Carlo (MC) simulations.[43,44]

## V. Acknowledgements

Work at Sun Yat-Sen University was supported by the National Natural Science Foundation of China (Grants No. 11904416, No. 12174454, No. 11904414, No. 12104518, No. 22090041, No. 92165204, No. 12074426, and No. 11227906), Guangdong Basic and Applied Basic Research Funds (Grant No. 2021B1515120015), National Key Research and Development Program of China (Grant No. 2019YFA0705702), the Fundamental Research Funds for the Central Universities and the Research

Funds of Renmin University of China (Grant No. 22XNKJ40). The work at Rice University is supported by the US Department of Energy, Basic Energy Sciences, under grant no. DE-SC0012311 and by the Robert A. Welch Foundation grant no. C-1839 (P.D.). We appreciate the support of BSRF, IHEP, CAS for high pressure XRD measurements.

1. Bardeen, J., Cooper, L. & Schrieffer, J. Microscopic Theory of Superconductivity. *Phys. Rev.* **106**, 162–164 (1957).
2. Si, Q., Yu, R. & Abrahams, E. High-temperature superconductivity in iron pnictides and chalcogenides. *Nat. Rev. Mater.* **1**, 16017 (2016).
3. Scalapino, D. J. A common thread : The pairing interaction for unconventional superconductors. *Rev. Mod. Phys.* **84**, 1383 (2012).
4. Fernandes, R. M. *et al.* Iron pnictides and chalcogenides : a new paradigm for superconductivity. *Nature* **601**, 35–44 (2022).
5. Stewart, G. R. Unconventional superconductivity. *Adv. Phys.* **66**, 75–196 (2017).
6. Dai, P. Antiferromagnetic order and spin dynamics in iron-based superconductors. *Rev. Mod. Phys.* **87**, 855–896 (2015).
7. Ran, S. *et al.* Extreme magnetic field-boosted superconductivity. *Nat. Phys.* **15**, 1250–1255 (2019).
8. Ran, S. *et al.* Nearly ferromagnetic spin-triplet superconductivity. *Science (80-. ).* **365**, 684–687 (2019).
9. Gammel, P. C. C. L., Bishop, D. J., Canfield, P. C., Gammel, P. L. & Bishop, D. J. New magnetic Superconductors : A Toy Box For Solid-State Physicists. *Phys. Today* **51**, 40–46 (1998).
10. Yu, J., Le, C., Li, Z. & Li, L. Coexistence of ferromagnetism , antiferromagnetism , and superconductivity in magnetically anisotropic (Eu,La)FeAs2. *npj Quantum Mater.* **6**, 63 (2021).
11. W. T. Jin, Y. Xiao, Z. Bukowski, Y. Su, S. Nandi, A. P. Sazonov, M. Meven, O. Zaharko, S. Demirdis, K. N. Phase diagram of Eu magnetic ordering in Sn-flux-grown Eu(Fe1−xCox)2As2 single crystals. *Phys. Rev. B* **94**, 184513 (2016).
12. Jaccarino, V. & Peter, M. Ultra-High-Field superconductivity. *Phys. Rev. Lett.* **9**, 290 (1962).
13. Briggs, A. Observation of Magnetic-Field-Induced Superconductivitity. *Phys. Rev. Lett.* **53**, 497 (1984).
14. Yin, J. *et al.* Large negative magnetoresistance in the antiferromagnetic rare-earth dichalcogenide EuTe2. *Phys. Rev. Mater.* **4**, 13405 (2020).
15. Yang, H. *et al.* Colossal angular magnetoresistance in the antiferromagnetic semiconductor EuTe2. *Phys. Rev. B* **104**, 214419 (2021).
16. Pan, X. C. *et al.* Pressure-driven dome-shaped superconductivity and electronic structural evolution in tungsten ditelluride. *Nat. Commun.* **6**, 7805 (2015).
17. Cai, W. *et al.* Pressure-induced superconductivity and structural transition in ferromagnetic CrSiTe3. *Phys. Rev. B* **102**, 144525 (2020).
18. Yu, F. H. *et al.* Elevating the magnetic exchange coupling in the compressed antiferromagnetic axion insulator candidate EuIn2As2. *Phys. Rev. B* **102**, 180404 (2020).
19. Sun, H. *et al.* Magnetism variation of the compressed antiferromagnetic topological insulator EuSn2As2. *Sci. China Physics, Mech. Astron.* **64**, 118211 (2021).
20. Yang, P. T. *et al.* Pressured-induced superconducting phase with large upper critical field and concomitant enhancement of antiferromagnetic transition in EuTe2. *Nat. Commun.* **13**, 2975 (2022).
21. Wang, Y. *et al.* An effective structure prediction method for layered materials based on 2D particle swarm optimization algorithm. *J. Chem. Phys.* **137**, 224108 (2012).
22. Wang, Y., Lv, J., Zhu, L. & Ma, Y. Crystal structure prediction via particle-swarm optimization. *Phys. Rev. B* **82**, 094116 (2010).

23. Wang, Y., Lv, J., Zhu, L. & Ma, Y. CALYPSO: A method for crystal structure prediction. *Comput. Phys. Commun.* **183**, 2063–2070 (2012).

24. Coelho, A. A. TOPAS and TOPAS-Academic : an optimization program integrating computer algebra and crystallographic objects written in C++. *J. Appl. Crystallogr.* **51**, 210 (2018).

25. Ohara, H. *et al.* Charge ordering in $Eu_3S_4$ determined by the valence-difference contrast of synchrotron X-ray diffraction. *Phys. B Condens. Matter* **350**, 353–365 (2004).

26. Search, H., Journals, C., Contact, A. & Iopscience, M. Bcc-fcc structure transition of Te. *J. Phys. Conf. Ser.* **500**, 192018 (2014).

27. Jamieson, J. C. & Mcwhan, D. B. Crystal Structure of Tellurium at High Pressures. *Phys. Today* **43**, 1149 (2014).

28. Bacon, J., Sciences, B. & Bnj, B. Induction of superconductivity by applied magnetic fields. *Nature* **315**, 95 (1985).

29. Clogston, A. M. Upper Limit for the Critical Field in Hard Superconductors. *Phys. Rev. Lett.* **9**, 266–267 (1962).

30. Akiba, K. *et al.* Magnetotransport properties of tellurium under extreme conditions. *Phys. Rev. B* **101**, 245111 (2020).

31. Qi, Y. *et al.* Pressure-driven superconductivity in the transition-metal pentatelluride $HfTe_5$. *Phys. Rev. B* **94**, 054517 (2016).

32. Zhang, J. L. *et al.* Pressure-induced superconductivity in topological parent compound $Bi_2Te_3$. *Proc. Natl. Acad. Sci. U. S. A.* **108**, 24–28 (2011).

33. Malliakas, C. D., Chung, D. Y., Claus, H. & Kanatzidis, M. G. Superconductivity in the Narrow-Gap Semiconductor $CsBi_4Te_6$. *J. Am. Chem. Soc.* **135**, 14540 (2013).

34. Butch, N. P. Expansion of the high fi eld-boosted superconductivity in $UTe_2$ under pressure. *npj Quantum Mater.* **6**, 75 (2021).

35. Jiao, L. *et al.* Chiral superconductivity in heavy-fermion metal $UTe_2$. *Nature* **579**, 523 (2020).

36. Duan, C. *et al.* Resonance from antiferromagnetic spin fluctuations for superconductivity in $UTe_2$. *Nature* **600**, 636 (2021).

37. Cheng, P. *et al.* Nuclear Instruments and Methods in Physics Research A Design of the cold neutron triple-axis spectrometer at the China Advanced Research Reactor. *Nucl. Inst. Methods Phys. Res. A* **821**, 17–22 (2016).

38. Prescher, C. & Prakapenka, V. B. DIOPTAS : a program for reduction of two- dimensional X-ray diffraction data and data exploration. *High Press. Res.* **35**, 223–230 (2015).

39. Perdew, J. P., Burke, K. & Ernzerhof, M. Generalized gradient approximation made simple. *Phys. Rev. Lett.* **77**, 3865–3868 (1996).

40. Joubert, D. From ultrasoft pseudopotentials to the projector augmented-wave method. *Phys. Rev. B* **59**, 1758 (1999).

41. Kresse, G. & Furthmüller, J. Efficiency of ab-initio total energy calculations for metals and semiconductors using a plane-wave basis set. *Comput. Mater. Sci.* **6**, 15–50 (1996).

42. Vargas-Hernández, R. A. Bayesian Optimization for Calibrating and Selecting Hybrid-Density Functional Models. *J. Phys. Chem. A* **124**, 4053–4061 (2020).

43. Lou, F. *et al.* PASP: Property analysis and simulation package for materials. *J. Chem. Phys.* **154**, 114103 (2021).

44. Hukushima, K. & Nemoto, K. Exchange Monte Carlo Method and Application to Spin Glass Simulations. *Journal of the Physical Society of Japan* **65**, 1604–1608 (1996).

# Supplementary: Exchange field enhanced upper critical field of the superconductivity in compressed antiferromagnetic EuTe$_2$

Hualei Sun[1], Liang Qiu[1], Yifeng Han[2], Yunwei Zhang[1], Weiliang Wang[3], Chaoxin Huang[1], Naitian Liu[1], Mengwu Huo[1], Lisi Li[1], Hui Liu[1], Zengjia Liu[1], Peng Cheng[4], Hongxia Zhang[4], Hongliang Wang[5], Lijie Hao[5], Man-Rong Li[2], Dao-Xin Yao[1], Yusheng Hou[1], Pengcheng Dai[4], and Meng Wang[1,*]

[1]*Center for Neutron Science and Technology, Guangdong Provincial Key Laboratory of Magnetoelectric Physics and Devices, School of Physics, Sun Yat-Sen University, Guangzhou, Guangdong 510275, China*

[2]*Key Laboratory of Bioinorganic and Synthetic Chemistry of Ministry of Education, School of Chemistry, Sun Yat-Sen University, Guangzhou, Guangdong 510275, China*

[3]*School of Physics, Guangdong Province Key Laboratory of Display Material and Technology, Sun Yat-Sen University, Guangzhou, Guangdong 510275, China*

[4]*Laboratory for Neutron Scattering and Beijing Key Laboratory of Optoelectronic Functional Materials and MicroNano Devices, Department of Physics, Renmin University of China, Beijing 100872, China*

[5]*China Institute of Atomic Energy, PO Box-275-30, Beijing 102413, China*

[6]*Department of Physics and Astronomy, Rice Center for Quantum Materials, Rice University, Houston, TX, USA*

* wangmeng5@mail.sysu.edu.cn

## Structural transition

EuTe$_2$ undergoes an obviously structural transition at ~16.0 GPa. In the LP phase, the *in situ* high pressure XRD patterns can be well indexed by the trigonal *I*4/*mcm* space group. The XRD patterns of the HP-phase can be described by the calculated structure that belongs to the *C*2/*m* space group. The refinements for the XRD patterns at two selected pressures that correspond to the LP and HP phases are shown in Fig. S1. The related structural parameters are listed in Table S1.

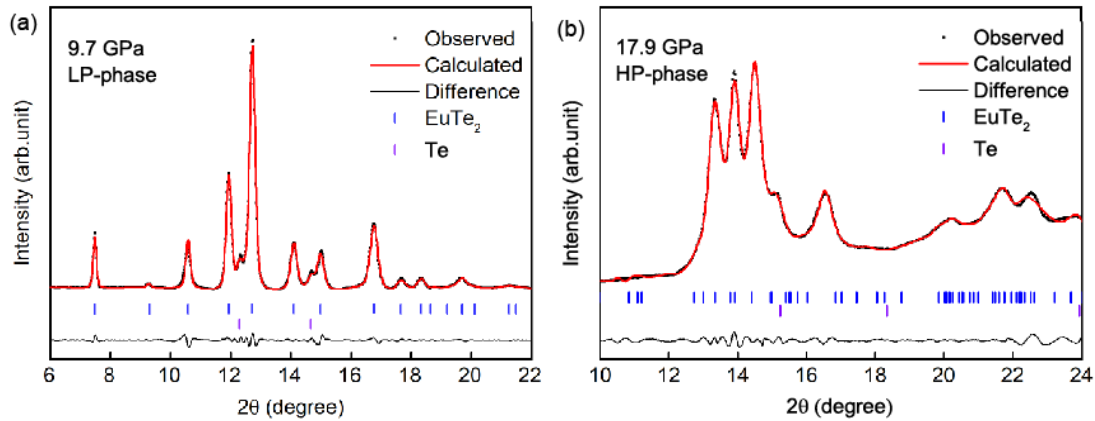

**Figure S1** (a) The refined XRD patterns of EuTe$_2$ at 9.7 GPa and (b) 17.9 GPa. The blue and purple vertical ticks mark the Bragg reflections of EuTe$_2$ and Te, respectively.

**Table S1** Refined lattice parameters, atomic coordinates, and the Wyckoff positions (WP) of EuTe$_2$ at 9.7 and 17.9 GPa.

| The LP phase, pressure at 9.7 GPa, Space group: *I*4/*mcm* |
|---|
| $a$ = 6.721(8), $c$ = 7.652(9), $R_{wp}$ = 4.67%, $R_p$ = 7.79% |

| atom | x | y | z | Occ. | WP |
|------|---|---|---|------|-----|
| Eu | 0.5 | 0.5 | 0.75 | 1 | 4a |
| Te | 0.1461(5) | 0.3539(5) | 0.5 | 1 | 8h |

The HP phase, pressure at 17.9 GPa, Space group: $C2/m$

$a = 10.451(13)$, $b = 3.524(5)$, $c = 9.544(11)$, $\beta = 131.68(4)°$, $R_{wp} = 5.12\%$, $R_p = 4.71\%$

| atom | x | y | z | Occ | WP |
|------|---|---|---|-----|-----|
| Eu | 1 | 0 | 0.835(5) | 1 | 4i |
| Te 1 | 0.25(4) | 0 | 0.83(3) | 1 | 4i |
| Te 2 | 0.86(4) | 0.5 | 0.52(3) | 1 | 4i |

## Hall resistance under pressure

The Hall resistance, $R_{xy}$, was measured from -10 to +10 T at pressures ranging from 2.3 to 27.0 GPa at 10 K. The results are displayed in Fig. S2(a). For the magnetic fields from 2 to 10 T, $(R_{xy}^+ - R_{xy}^-)/2$ vary approximately linearly versus the magnetic field. The Hall coefficient and carrier density are calculated from the slope of $(R_{xy}^+ - R_{xy}^-)/2$ and displayed in Fig. S2(b). The Hall coefficient remains positive, suggesting that the majority carriers are holes. The Hall coefficient decreases monotonically up to 15.0 GPa then remains constantly, resulting in a transition-like change between 15.0 and 18.0 GPa for the density of carriers. The density of carriers is $3.6\times10^{19}$cm$^{-3}$ at 2.3 GPa and $3.2\times10^{23}$cm$^{-3}$ at 18.0 GPa.

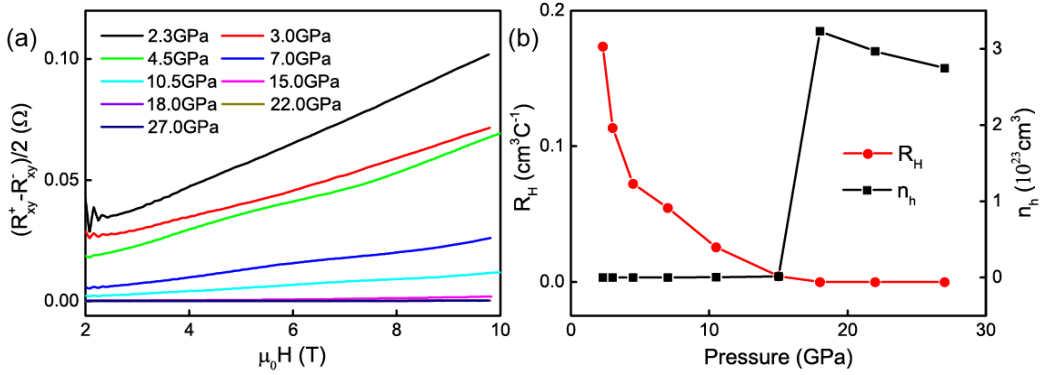

**Figure S2** (a) Hall resistance $(R_{xy}^+ - R_{xy}^-)/2$ at 10 K under various pressures. The $R_{xy}^+$ and $R_{xy}^-$ were measured with a positive and negative magnetic field, respectively. The direction of the magnetic field is along the $c$ axis. (b) Pressure dependence of Hall coefficients (red) and carrier densities (black) measured at 10 K. The Hall coefficients and carrier densities are calculated within the magnetic field in (a), where $(R_{xy}^+ - R_{xy}^-)/2$ is approximately linear in relation to the magnetic field.

## Magnetoresistance under pressure

The *MR*s under various pressures from 10 to 150 K are presented in Fig. S3. In this temperature range, EuTe$_2$ undergoes from an AFM state to a PM state. We define *MR* as $MR = \frac{\rho(H)-\rho(0)}{\rho(0)} \times 100\%$, where $\rho(H)$ and $\rho(0)$ are the resistance measured at magnetic field $\mu_0H$ and zero field, respectively. The *MR*s exhibit a significant reduction above 16.2 GPa compared with that of the low pressures at 3.6, 7.8, and 12.7 GPa. The similarities between the *MR*s in the PM state of the LP phase at 150 K and the HP phase at 10 K suggest that the HP phase is in a PM state.

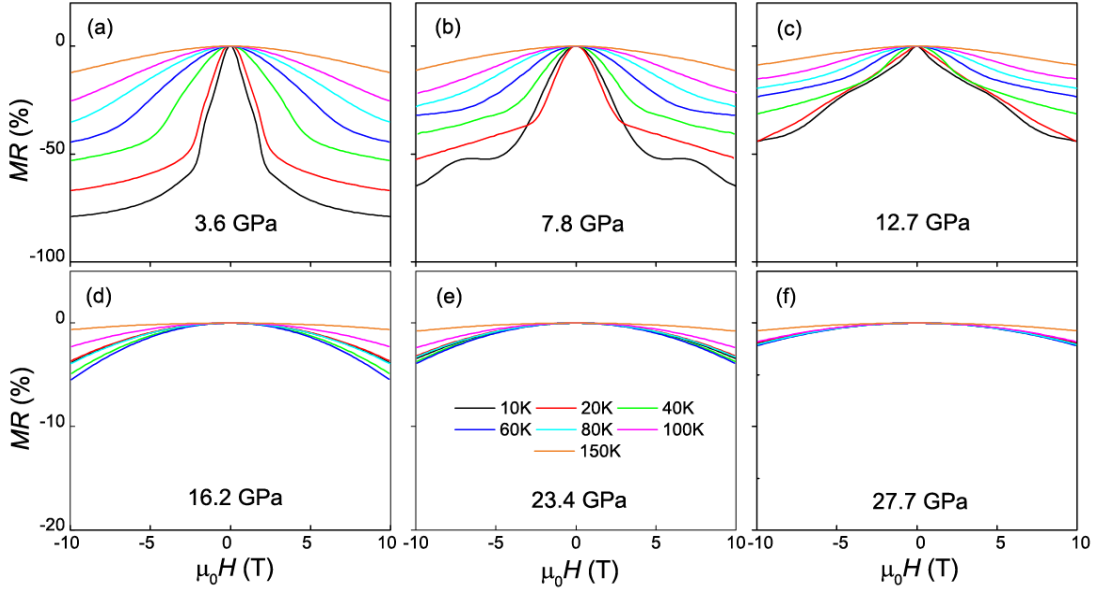

**Figure S3** (a)–(f) Magnetoresistance (*MR*) of EuTe$_2$ single crystals measured from 10 to 150 K at 3.6, 7.8, 12.7, 16.2, 23.4, and 27.7 GPa.

## High pressure SC transition

The low temperature resistance under various pressures in the LP-phase and HP-phase from different measurements is shown in Figs. S4(a) and S4(b), respectively. The resistance measured at 4.8 GPa shows a drop at 3.0 K. At 5.7 GPa, the resistance drops at 3.6 K and decreases to zero in the SC state at 2 K. The $T_c$ reaches to 6.1 K at 7.0 GPa and shows a slowly decreasing trend for higher pressures. When the crystal structure transforms into the HP phase, the $T_c$s show slightly declining with further increasing pressure and keep around ~5.5 K from 16.2 to 27.7 GPa.

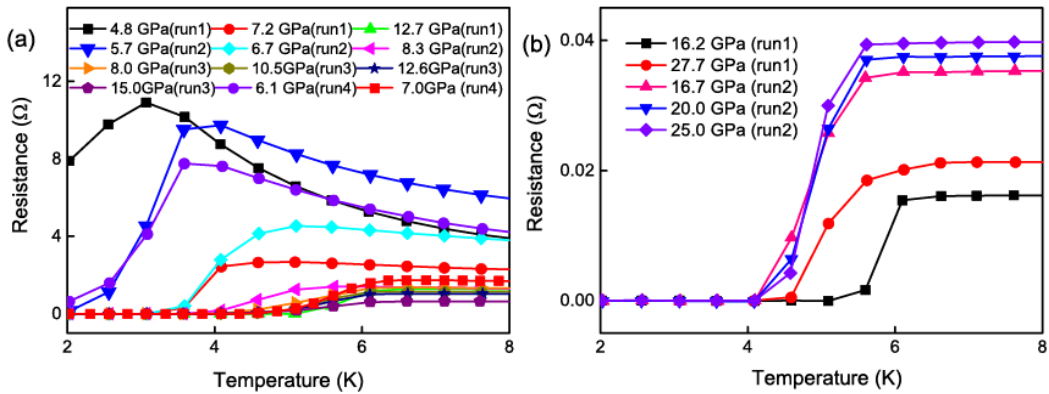

**Figure S4** (a) Low temperature resistance under various pressures of the LP-phase and (b) the HP-phase collected from different measurements (run 1 to run 4). The $T_c$s are determined by the intersections of the tangent to the resistance curve during the transition process and the straight-line of the normal state above the transition.

## Neutron diffraction measurements and the *C*-type AFM order

To determine the magnetic order, we conducted neutron diffraction experiment on Xingzhi triple-axis spectrometer at the China Advanced Research Reactor (CARR). Eu has a large neutron absorption cross section. Before experiment, we simulated the diffraction patterns for the *C*-type, *G*-type, and the *A*-type

AFM orders as shown in Figs. S5(a) and S5(b). We focused on the $2\theta$ angles of the calculations. Figure S5(c) shows the neutron diffraction pattern at 3.5 and 25 K. The data demonstrate EuTe$_2$ exhibits the *C*-type AFM at low temperature.

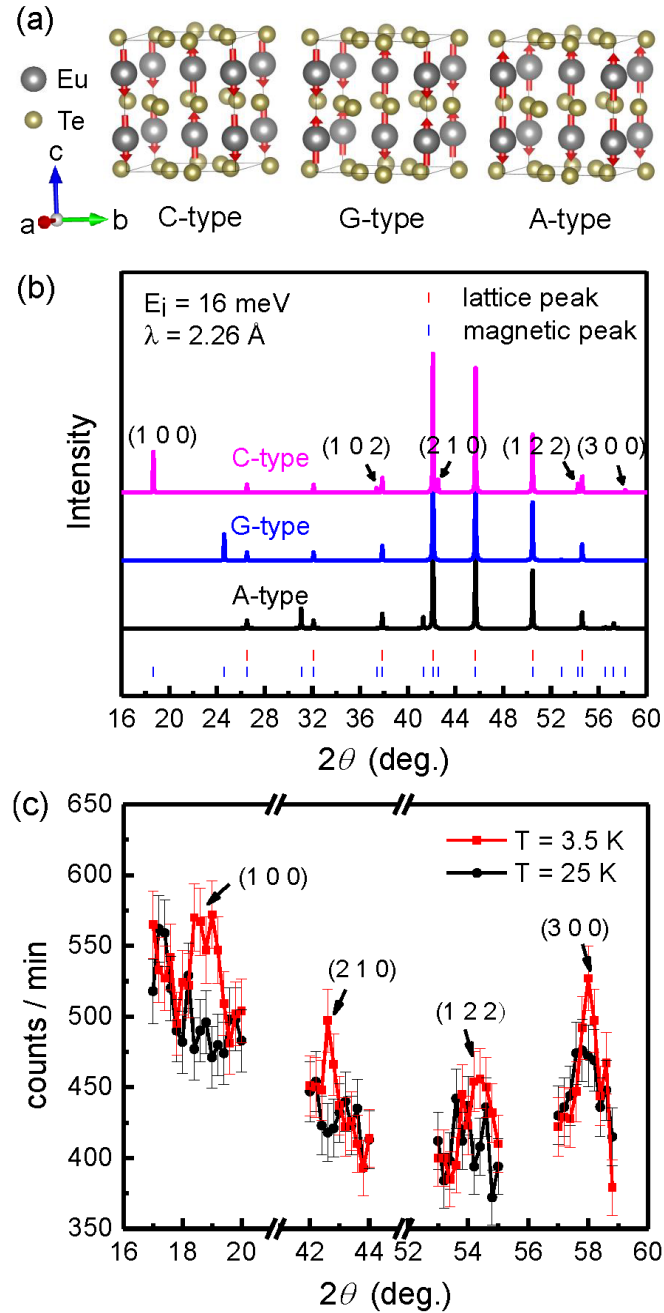

**Figure S5** (a) Three possible AFM configurations of EuTe$_2$. (b) Simulated powder neutron diffraction spectra for three different AFM configurations with neutron incident energy of $E_i$ = 16 meV ($\lambda$ = 2.26 Å), where the magnetic peaks of the C-type order are marked seriatim. The spectra have been vertically shifted for comparison. The positions of the lattice and magnetic peaks are marked by red and blue vertical ticks, respectively. (c) Powder neutron diffraction experimental results of EuTe$_2$ with the same neutron incident energy and wavelength as simulation. Two independent scans are performed below and above $T_N$ at $T$ = 3.5 and 25 K. The magnetic peaks at $Q$ = (1 0 0), (2 1 0), (1 2 2), and (3 0 0) of the *C*-type AFM can be observed clearly. The errors correspond to one standard deviation.

## *C*-type AFM transition under pressure

The AFM transition temperature $T_N$ is enhanced under pressure due to the increase of the magnetic exchange couplings. The $T_N$ could be determined from the kink on resistance of the LP phase. Figure S6 displays selected temperature dependence of resistance under various pressures from 0.8 to 8.0 GPa. The resistance curves are collected from different measurements, including run 1 to run 4. The kinks indicating the $T_N$s have been marked in these plots. The $T_N$s increase from 11.4 K at ambient pressure to 12.6 K at 1.5 GPa, and to 16.7 K at 8.0 GPa. For higher pressures in the LP phase, the signal of the AFM transition becomes weaker and cannot be identified from resistance. Combining the DFT calculations and Monte Carlo simulations, six-nearest neighbor exchange couplings are calculated. The $T_N$s are simulated accordingly. The results are listed in Table S2.

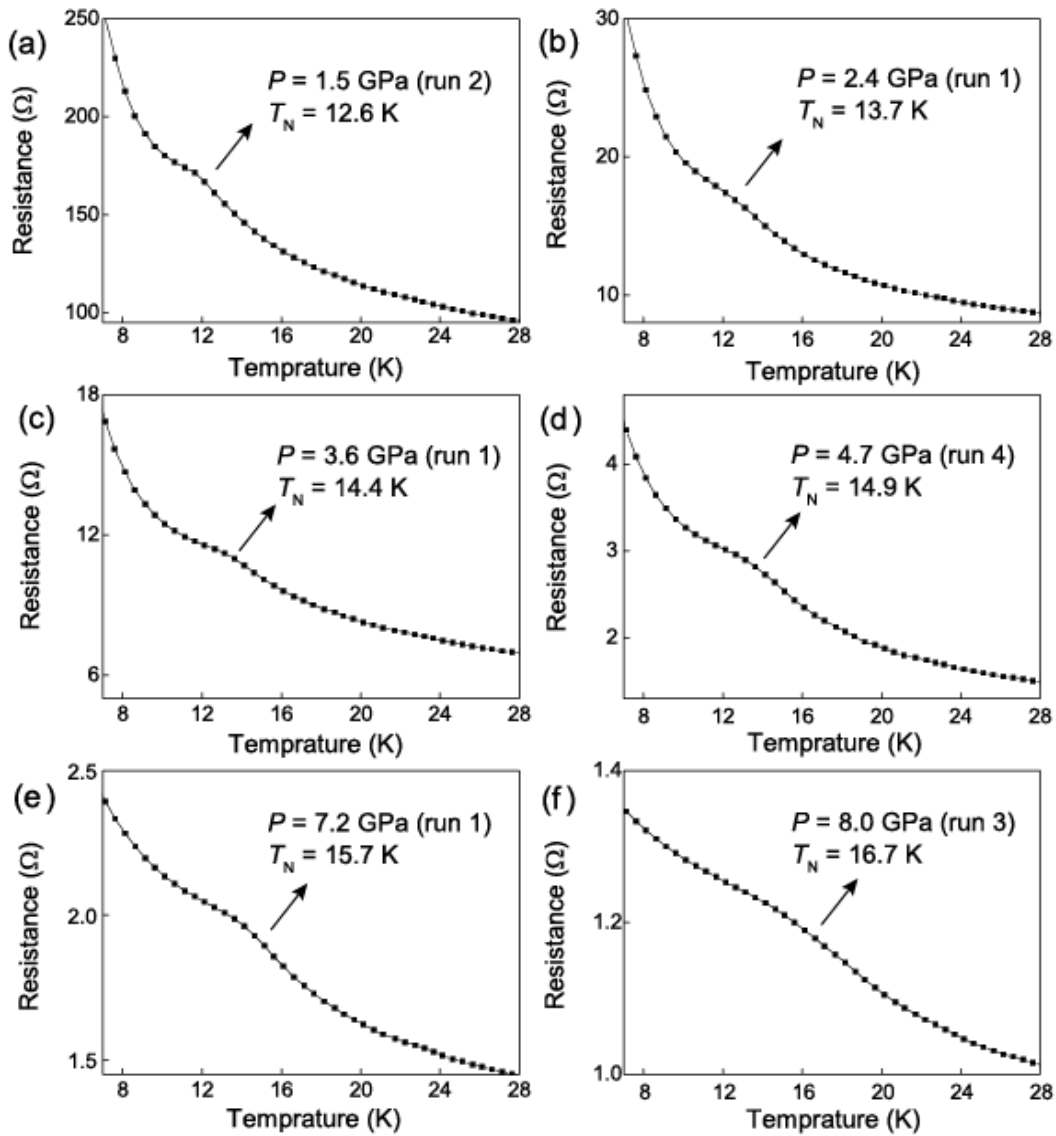

**Figure S6** Temperature dependent resistance under pressures of (a) 1.5, (b) 2.4, (c) 3.6, (d) 4.7, (e) 7.2, and (f) 8.0 GPa. The resistance curves are selected from different measurements (run 1 to 4). The $T_N$s are determined by the inflection points marked on the plots.

Table S2 Magnetic exchange couplings $J$s, $T_N$ and moment per $Eu^{2+}$ of $EuTe_2$ under pressure.

| Pressure(GPa) | $J_1$(meV) | $J_2$(meV) | $J_3$(meV) | $J_4$(meV) | $J_5$(meV) | $J_6$(meV) | $T_N$(K) | Moment($\mu_B$) |
|---|---|---|---|---|---|---|---|---|
| 0 | -0.905 | -0.065 | 0.422 | 0.092 | -0.012 | 0.015 | 13.17 | 6.943 |
| 2 | -1.031 | -0.082 | 0.448 | 0.115 | -0.034 | 0.027 | 14.63 | 6.941 |
| 4 | -1.332 | -0.117 | 0.468 | 0.132 | -0.036 | 0.005 | 16.09 | 6.94 |
| 5.9 | -1.670 | -0.140 | 0.483 | 0.175 | -0.016 | -0.024 | 16.29 | 6.938 |
| 7.8 | -1.767 | -0.169 | 0.493 | 0.222 | 0.008 | -0.066 | 17.56 | 6.937 |
| 9.7 | -1.994 | -0.197 | 0.509 | 0.195 | -0.028 | -0.104 | 20.48 | 6.936 |
| 11.8 | -2.387 | -0.229 | 0.527 | 0.243 | -0.037 | -0.109 | 21.21 | 6.935 |
| 14.3 | -3.232 | -0.173 | 0.483 | 0.358 | -0.043 | -0.035 | 16.10 | 6.934 |
| 15.9 | -3.658 | -0.180 | 0.506 | 0.407 | -0.092 | -0.023 | 15.36 | 6.933 |

## Spin flop and spin flip transitions at 7.0 GPa

At 7.0 GPa, the $T_N$ is increased to 15.6 K. The spin flop and spin flip transitions are expected to occur at higher magnetic fields compared to that at ambient pressure. The resistance as a function of magnetic field up to 14.0 T at various temperatures are shown in Fig. S7. The local magnetic moments are influenced by the external magnetic field and undergo the spin flop and spin flip transitions. The conduction electrons have correlations with the local magnetic moments, resulting in step-like kinks on resistance. Thus, the temperatures for the spin flop and spin flip transitions are determined in Figs. S7 (b-f). In the SC state, the electrons that form cooper pairs do not interact with the local moments. The spin flop transition at 3.75 K in Fig. S7 (a) which should be at ~5.5 T could not be identified.

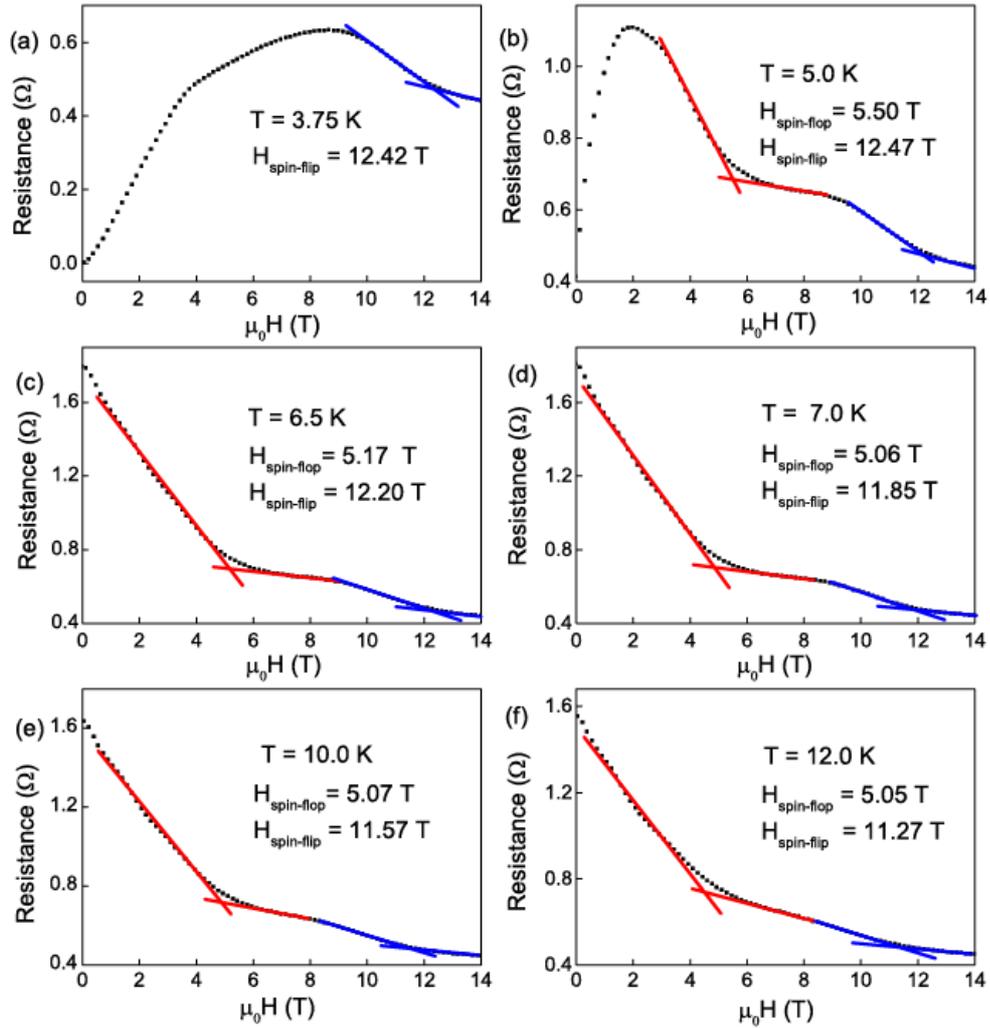

**Figure S7** Magnetoresistance curves at various temperatures of (a) 3.75, (b) 5.0, (c) 6.5, (d) 7.0, (e) 10.0, and (f) 12.0 K. The spin flop and spin flip transition temperatures are determined by the intersections of the tangents to the resistance curves during the decline and the stable region.

## Ginzburg-Landau fitting of the $H_{c2}$

As the Jaccarino-Peter mechanism, a net FM magnetization will produce an exchange field and result in a change of the upper critical field $H_{c2}$ for a superconductor. We indeed observe a $H_{c2} - T_c$ relation that deviates from a simple Ginzburg-Landau (GL) formula as shown in Fig. S8. The $H_{c2}$s are separated into three segments as the distinct magnetic sates: the AFM, spin flop, and spin flipped states. The $H_{c2} - T_c$ relation in each segment is fitted to the GL formula separately. The $T_c$, $\rho_0 H_{c2}$, and spin state have been clarified in Fig. S8.

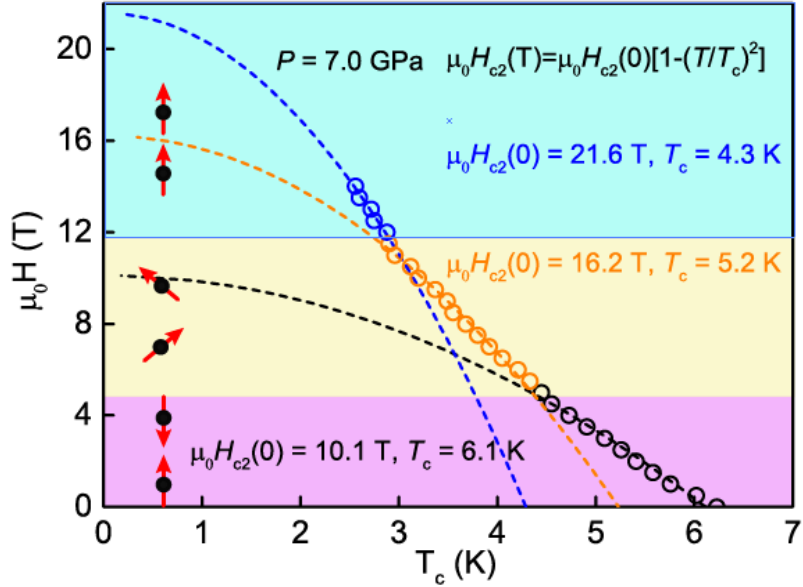

**Figure S8** GL fittings of the $\mu_0H_{c2}$ based on the $T_c$s determined from temperature dependent resistance. Different colors represent distinct magnetic moment textures of the $Eu^{2+}$ spins. The Pink, yellow, and blue regions correspond to the AFM, spin flop, and spin flipped state, respectively. The black, orange, and blue dashed lines represent three GL fittings.

## Band structure calculations

The band structures of the SC state in the LP phase are calculated by the density functional theory. The results reveal that the density of states on the Fermi surface is contributed by the Te $5p$ electrons. The Eu $4f$ electrons are localized around $-1.25$ eV below the Fermi level.

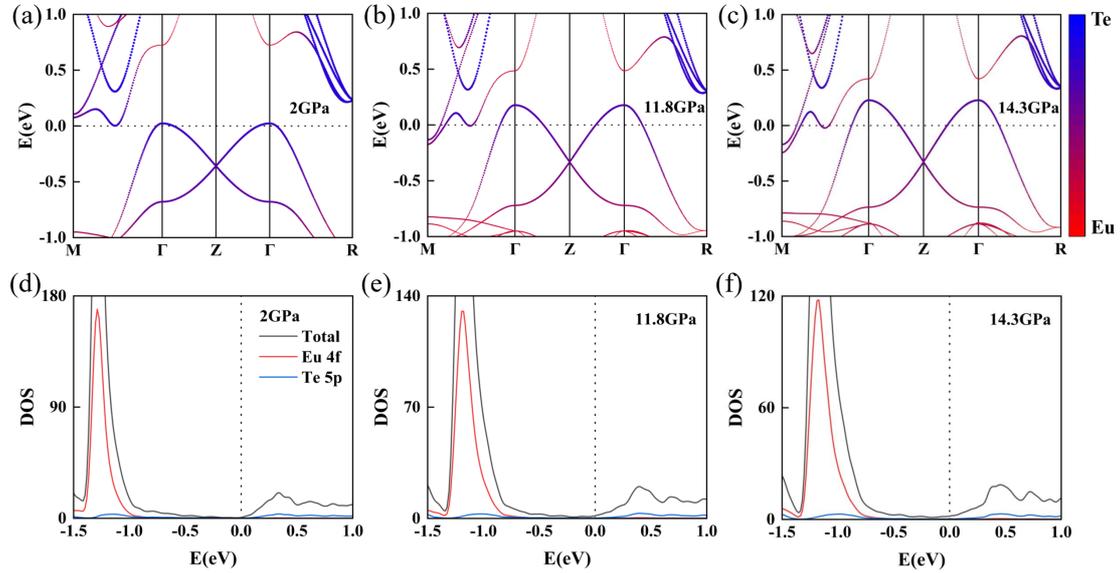

**Figure S9** Band structures and density of states (DOS) of $EuTe_2$ under different pressures with spin orbital coupling.